\documentclass{lncse}
\def\Journal#1#2#3#4{{#1} {\bf #2}, #3 (#4)}

\def\NPB{{\em Nucl. Phys.} B}
\def\PLB{{\em Phys. Lett.}  B}
\def\PRL{\em Phys. Rev. Lett.}
\def\PRD{{\em Phys. Rev.} D}

\def\IJMP{{\em Int. J. Mod. Phys.} C}
\begin{document}
\title{{\hfill \normalsize RUHN-99-3}\\
The Overlap Dirac Operator
\thanks{
Talk delivered at the ``Interdisciplinary Workshop
on Numerical Challenges to Lattice QCD'', Wuppertal, August 22-24, 1999.}
}
\titlerunning{Overlap Dirac Operator}
\author{Herbert Neuberger}
\authorrunning{Neuberger}
\institute{Rutgers University\\Department of Physics and Astronomy\\
Piscataway, NJ08855, USA.}
\maketitle
\begin{abstract}
This introductory presentation describes the Overlap Dirac Operator,
why it could be useful in numerical QCD, and how it can be implemented.
\end{abstract}
\section{Introduction}
The objective of this talk is to introduce a certain matrix, the Overlap
Dirac Operator. The ideas behind this construction are rather deep, 
having to do with the self-consistency of chiral gauge theories
\cite{overlap}. Chiral
gauge theories are a class of quantum field theoretical models from
which one draws the minimal standard model of particle physics. This
model describes all phenomena at distances larger than about 
$10^{-18}$ meters, excluding gravity. 

The Overlap Dirac Operator
may be useful also in numerical studies of the strong interaction
component of the minimal standard model, Quantum Chromo-Dynamics
(QCD). QCD, in isolation, is  a vector-like gauge theory, a simple 
combination of chiral gauge theories within which some of the 
mathematical complexities of the general class disappear. Nevertheless,
the strong nonlinearity of QCD makes the quantitative evaluation
of its most basic observable features possible only within large scale
numerical Monte Carlo simulations. This workshop will deal 
mainly with numerical QCD, the subfield of lattice field theory
which focuses on such simulations.  

Therefore, the form of the Overlap Dirac Operator will be motivated 
heuristically, based on technical considerations of numerical 
QCD. The heuristic motivation will start from the continuum Euclidean 
Dirac operator, whose basic features of direct relevance to numerics 
will be reviewed. As alluded above, 
this is not the original way the Overlap Dirac
Operator was found, but, for the focus
of this workshop it is unnecessary to go through the entire construction.
Rather, I shall start by reviewing an important technicality, 
the light quark bottleneck, which is a serious 
problem in contemporary traditional numerical QCD, 
argue that there is no fundamental
reason why this bottleneck cannot be avoided and present the Overlap Dirac 
Operator as a potential solution\footnote{Other
solutions than the Overlap Dirac Operator have been proposed for this 
specific problem; using the Overlap Dirac Operator is probably 
the most fundamentally motivated approach, but the cost may be high,
and at the bottom line this consideration often takes precedence.}. 
The solution is not yet working efficiently
because another bottleneck appears. However, the new bottleneck may
be easier to circumvent than the old one. 

An effort has been made
to make this presentation accessible to applied mathematicians. 
Physicists practicing numerical QCD, in particular when the Overlap
Dirac Operator is used, are facing issues that the expertise of specialists 
in numerical linear algebra could be of great help in resolving.
These same physicists will likely be familiar
with large portions of the presentation
below, perhaps excepting the way the Overlap Dirac
Operator is arrived at. 
\section{The numerical bottleneck of light quarks}
The QCD Lagrangian 
with correct parameters should produce the observed ratio of masses
${m_\pi \over m_\rho} =0.18$. The relative smallness of this number
reflects the relative small mass of two of the lightest quarks in
Nature. This mass is not only small relatively to other quark masses,
but, more importantly, relative to a basic scale of QCD itself, a mass
parameter known as $\Lambda_{QCD}$. Theoretically, one has very
good reasons to expect to be able to smoothly extrapolate in the light
quark masses from, say, ${m_\pi \over m_\rho} =0.25$ to the smaller
physical value. Actually, even the functional form of these extrapolations is
known. However, it is not likely that the extrapolation 
will go smoothly through the value ${m_\pi \over m_\rho} =0.5$.
As long as ${m_\pi \over m_\rho} \ge 0.5$ the decay  $\rho \to 2\pi$
is energetically forbidden but once this threshold is passed the decay becomes
allowed. This provides physical reasons to expect 
a ``bumpy'' behavior\footnote{Some quantities may be insensitive
to the  $\rho \to 2\pi$ threshold effects, 
typically for kinematic reasons. Also, there is an additional
approximation which is very often employed (the so called ``valence'',
or ``quenched'' approximation) which eliminates decay effects
on the vacuum of the theory. On the other hand, one often relies
on an effective low energy description of $QCD$, in which the 
Lagrangian is replaced by an ``effective Lagrangian'' involving only
$\pi$'s and some other light associated particles. The existence
of a particle like the $\rho$ and its connection to the $2\pi$ state
is then encoded in the coefficients of the ``effective Lagrangian''
and will affect the accuracy of chiral perturbation theory at
moderate energies. Thus,
physical observables related to ``weak matrix
elements'' are an example where one might suspect sensitivity 
to threshold effects.}
in the extrapolation to light quark masses in
the region where ${m_\pi \over m_\rho} \sim 0.5$ and there are no first
principle based 
theoretical expressions parameterizing this ``bumpy'' region.  
Current large scale numerical
work happens to be 
close to the threshold ${m_\pi \over m_\rho} =0.5$, and there
are numerical difficulties obstructing straightforward attempts
to go significantly lower. 

In numerical work
one uses a lattice whose spacing $a$ 
is measured in inverse mass units. Thus any number of the type mass $\times$
$a$ is a pure number, of the kind a computer can handle. The threshold
ratio is obtained, with current simulation parameters, for $m_\pi a \sim 0.1$
and for quark mass $m_q$ given by $m_q a \sim 0.05$. 
Nothing is lost by setting $a=1$; dimensional considerations
can always restore the right power of $a$. Physical
considerations imply $m_\pi \propto \sqrt {m_q}$ for small 
$m_q$. On the other hand,  
$m_\rho$ stays finite (basically of the  
order of $\Lambda_{QCD}$) at $m_q =0$ and consequently 
its numerical dependence 
on $m_q$ for small $m_q$ is weaker. So, in order to reduce
${m_\pi \over m_\rho}$ 
by a factor of 2 as we would like, we have to reduce the light quark
masses by about a factor of 4. The bottom line is that we would
like to leave all other simulation parameters the same, only
change the quark mass to something like $m_q a = 0.01$. 

Simple arguments indicate that we should be able to do that.
The quark mass enters as an important parameter in the lattice
Dirac Operator, $D$. $D$ and $D^\dagger D$ 
are matrices that need to be inverted
on vectors in the most inner procedure of the simulations. In
traditional simulations $D$ is a sparse matrix; full storage is
out of the question, as the dimension of $D$ is too large. 
The quark mass directly affects the condition number of $D$,
so its value can easily become the bottleneck of the entire
simulation. 
But, we could easily imagine being lucky enough
to manage the needed value of $m_q a = 0.01$: $D$ is a discretized
first order 
partial differential operator and, since space time is four
dimensional one can easily think of $D$ obeying $\| D \| \le 4$
as a uniform upper bound. 
On the other hand, $m_q$ enters
additively, with unit coefficient and, based on the continuum
formula one would expect that $D$ is 
approximately anti-hermitian at zero quark mass. Therefore, 
we expect $\| D^\dagger D \| \ge m_q^2$ and, 
when we invert $D^\dagger D$, 
a condition number of the order $10^5$, which should allow
the evaluation of ${1\over {D^\dagger D}}\psi$ in something like
500-1000 iterations. 

Up to relatively short lived jerks, 
computational power is projected to increase 
by Moore's Law for another decade, so we can expect an enhancement
by a factor of 250 at fixed power dissipation by 2010. Based on
the accumulated numerical 
experience to date, we can reasonably expect that, by then,
we would be able to carry out our computations with relative ease.

However, the sad fact is that theoretical issues have apparently made
it impossible to find a simple enough $D$ which would obey the 
rather plausible lower bound $\| D^\dagger D \| \ge m_q^2$. This
was a key assumption I made before in order to come up with the
condition number of order $10^5$. 
What really happens in
the traditional approach\footnote{To be specific, the approach
employing Wilson fermions; there also exists a traditional
``staggered fermion''
alternative, but it suffers from other problems.} 
is that $D^\dagger +D$ is not small and to get 
${m_\pi \over m_\rho}\sim .3$ one is forced into a regime where 
the lowest eigenvalue of $D^\dagger D$ quite often 
fluctuates as low as $10^{-8}$. This makes the condition 
number go to $10^{10}$ and
puts the ${m_\pi \over m_\rho}\sim .3$ mass ratio out of 
numerical reach. 

The above numerical problem reflects a difficulty of principle associated
with the $m_q =0$, or massless, case. In the continuum model, $m_q=0$
is associated with an extra symmetry, chirality. Technically, chirality
ensures that in the continuum the massive Dirac operator $D+m_q$ (where
I reserved $D$ for the massless case) indeed 
obeys  $\| D^\dagger D \| \ge m_q^2$. But, chirality cannot be preserved 
on the lattice in the same way as some other symmetries can. For twenty
years it was thought that an acceptable lattice version of $D$ 
which also preserves chirality does not exist. 
The good news is that a series of developments,
started in 92 by Kaplan and by Frolov and Slavnov \cite{kapslav}
and built on by
Narayanan and myself, have produced, in 97, the Overlap Dirac Operator
$D_o$ on the lattice, which, effectively has the extra symmetry. 
While $D_o$ is easy to write down, it is not easy to implement because
it no longer is sparse. There nevertheless 
is some hope, because, in the most common
implementation of the Overlap ideas, $D_o$ is what probably would be
the next best thing: up to trivially implemented factors,
it is a function (as an operator) of a sparse
matrix\footnote{The additional 
``trivial'' factors have a highly non-trivial numerical impact:
because of them the {\it inverse} of $D_o$ no longer has a simple
expression in terms of a function of a sparse matrix.}. 
Unfortunately, the function has a discontinuity, so its
implementation is costly. The numerical developments in this 
subfield are relatively
fresh and substantial progress has been achieved but 
we are not yet adequately tooled for full ``production runs''. 
Still, since the time we have had to address the problem
is of the order of a year or two I think further substantial progress
is likely. 
\section{Dirac Operator basics}
The Dirac Operator is a very important object in relativistic Field Theory.
Among other things, it predicted the existence of the electron anti-particle,
the positron. All of known matter is governed by the Dirac equation.  
\subsection{Continuum}
The Dirac Operator is defined in the continuum (Euclidean space) by
\begin{equation}
D_c =\sum_\mu \gamma_\mu (\partial_\mu + iA_\mu ) \enspace ,
\end{equation}
where,
\begin{equation}
\gamma_\mu^\dagger =\gamma_\mu ; ~~~\mu=1,2,3,4; 
~~~ \{ \gamma_\mu ,\gamma_\nu \}
\equiv \gamma_\mu \gamma_\nu + \gamma_\nu \gamma_\mu =2\delta_{\mu\nu}
\enspace .
\end{equation}
We also have,
\begin{equation}
A_\mu = A_\mu^\dagger ;~~~tr A_\mu =0 ; ~~~~ A_\mu {\rm ~~are~3~by~3~matrices.}
\end{equation}
The notation $\partial_\mu$ indicates $\partial /\partial x_\mu$. $A_\mu$
is $x$-dependent, but the $\gamma_\mu$ are constant four by four
matrices and operate in a separate space. Thus, $D_c$ is a twelve
by twelve matrix whose entries are first order partial differential
operators on a four dimensional Euclidean space, henceforth assumed
to be a flat four torus. The massive Dirac Operator is $m_c + D_c$.

The main properties of the Dirac Operator are:
\begin{enumerate}
\item  For $A_\mu =0$ one can set by Fourier transform 
$\partial_\mu = ip_\mu$ giving $D_c = i \gamma_\mu p_\mu$. This
produces $D_c^2 = -\sum_\mu p_\mu^2$ from the algebra of the 
$\gamma$-matrices. The important consequence this has is 
that $D_c=0$ iff $p_\mu =0$ for all $\mu$. 
\item  Define 
$\gamma_5 \equiv \gamma_1 \gamma_2 \gamma_3 \gamma_4$ which implies
$\gamma_5^2 =1$ and $\gamma_5 \gamma_\mu +\gamma_\mu \gamma_5 =0$. 
The property of $\gamma_5$-hermiticity is that
$D_c$ is hermitian under an indefinite inner product defined by
$\gamma_5$, or, equivalently, using the standard $L_2$ inner product,
we have for any backgrounds $A_\mu$ and any mass parameter $m_c$
\begin{equation}
\gamma_5 (D_c + m_c)\gamma_5 = (D_c + m_c )^\dagger \enspace .
\end{equation}
Thus, $\gamma_5$-hermiticity is a property of the massive Dirac
Operator. 
\item
At zero mass the Dirac Operator is anti-hermitian $D_c^\dagger = - D_c$ 
This property, in conjunction with $\gamma_5$-hermiticity implies the
symmetry property of chirality: $\gamma_5 D_c \gamma_5 = -D_c$ usually
written as $\{ D_c , \gamma_5 \} =0$. Traditionally, anti-hermiticity
would be viewed as a separate and trivial property and chirality
as a symmetry. Then, $\gamma_5$ hermiticity would be a consequence
holding not only for the massless Dirac Operator, but also
for non-zero mass. But, for better comparison to the lattice situation
it is somewhat better to exchange cause and consequence, as done here. 
\end{enumerate}

\subsection{Lattice}
Numerical QCD spends most of the machine cycles inverting a lattice
version of the massive Dirac Operator. This is how the continuum
Dirac operator is discretized:

The continuum $x$ is replaced by a discrete lattice site $x$ on a torus.
The matrix valued functions $A_\mu$ are replaced by 
unitary matrices $U_\mu (x)$
of unit determinant. $U_\mu (x)$ is associated with the link $l$ going
from $x$ into the positive $\mu$-direction, $\hat \mu$. 

The connection to the continuum is as follows: Assume that the 
functions\footnote{Actually, the $A_\mu(x)$ aren't really functions, 
rather they make up a  
one form $\sum_\mu A_\mu (x) dx_\mu$ which is a connection on a possibly
nontrivial bundle with structure group $SU(3)$ over the four-torus.
This complication is important for the case of massless quarks,
but shall be mostly ignored in the following.}
$A_\mu (x)$
are given. Then,
\begin{eqnarray}
U_\mu &= \lim_{N\to\infty} \left [ e^{iaA_\mu(x)/N}e^{iaA_\mu(x+a)/N} 
e^{iaA_\mu(x+2a)/N} \dots e^{iaA_\mu(x+(N-1)a))/N}\right ]
\nonumber \\ &\equiv
P \exp [i\int_l dx_\mu A_\mu (x) ]~~~~~{\rm (
the~symbol~P~denotes~``path~ordering" )}\enspace .
\end{eqnarray}

There are two sets of basic unitaries acting on twelve component
vectors $\psi(x)$:
\begin{enumerate}
\item The point wise acting $\gamma_\mu$'s and
\item the directional parallel transporters $T_\mu$, defined
by:
\begin{equation}
T_\mu (\psi ) (x) = U_\mu (x) \psi (x+\hat\mu) \enspace .
\end{equation}
\end{enumerate}
There also is a third class of unitaries carrying out gauge transformations.
Gauge transformations are characterized by a collection of $g(x)\in SU(3)$
and act on $\psi$ pointwise. The action is represented by the unitary
$G(g)$, so that $(G(g)\psi)(x) =g(x)\psi(x)$. Probably the most important
property of the $T_\mu$ operators is that they are ``gauge covariant'',
\begin{equation}
G(g) T_\mu (U) G^\dagger (g) = T_\mu (U^g ) \enspace ,
\end{equation}
where,
\begin{equation}
U^g_\mu (x) = g(x) U_\mu (x) g^\dagger (x+\hat\mu ) \enspace .
\end{equation}
The variables $U_\mu (x)$ are stochastic, distributed according to a
probability density that is invariant under $U\to U^g$ for any $g$. 

The lattice replacement of the massive continuum Dirac Operator, $D(m)$,
is an element in the algebra generated 
by $T_\mu,~ T_\mu^\dagger ,~ \gamma_\mu$.
Thus, $D(m)$ is gauge covariant. For $U_\mu (x) = 1$ the $T_\mu$ become
commuting shift operators. One can preserve several crucial 
properties of the continuum
Dirac Operator by choosing a $D(m)$ which satisfies:
\begin{enumerate}
\item Hypercubic symmetry. This discrete symmetry group consists of 24 
permutations scrambling the $\mu$ indices (which leave $D(m)$ unchanged)
combined with 16 reflections made out of the following four elementary
operations:
\begin{equation}
\gamma_5 \gamma_\mu D(T_\mu \longleftrightarrow T_\mu^\dagger )(m) \gamma_\mu
\gamma_5 = D(m) \enspace .
\end{equation} 
\item Correct low momentum dependence. When the 
commutators $[T_\mu , T_\nu ]=0$ one can simultaneously diagonalize the
$T_\mu$ unitaries, with eigenvalues $e^{i\theta_\mu^k}$. For small $\theta$
angles, 
one gets $D(m)\sim m + i\sum_\mu \gamma_\mu \theta_\mu + O(\theta^2 )$. One
also needs to require that 
for zero mass ($m=0$) $D^\dagger D$ be bounded away from zero for all
$\theta$ which are outside a neighborhood $R$ of zero. 
\item Locality and smoothness in the gauge field variables. 
One can also require, at least
for gauge backgrounds with small enough (in norm) commutators $[T_\mu,T_\nu]$,
that $D$ be a convergent series in the $T_\mu$.
\end{enumerate} 

The simplest solution to the above requirements is the Wilson Dirac Operator,
$D_W$. It is the sparsest possible realization of the above. 
Fixing a particular free parameter (usually denoted by $r$) 
to its preferred value ($r=1$)
$D_W$ can be written as:
\begin{equation}
D_W = m+4-\sum_\mu V_\mu ;~~~~V_\mu^\dagger V_\mu =1;~~~V_\mu=
{{1-\gamma_\mu}\over 2} T_\mu +{{1+\gamma_\mu}\over 2} T_\mu^\dagger
\enspace .
\end{equation}
For $m > 0$ clearly $\| D_W \| \ge m$, but to get the physical quark
mass to zero, $m_q^{\rm phys}=0$, one needs to take into account an
additive renormalization induced by the fluctuations in the gauge
background $U_\mu (x)$: $m$ must be chosen as $m=m_c (\beta ) < 0$
where $\beta$ is the lattice version of the gauge coupling constant
governing the fluctuations of the background. 
The need to use a negative $m$ on the lattice opens the door for very small
eigenvalues for $D_W^\dagger D_W$ and thus for terrible condition numbers.
To get a small but nonzero physical quark mass one should choose
$m=m_c(\beta ) + \Delta m,~\Delta m > 0$ but the numbers are such
that one always ends up with an overall negative $m$.

\subsection{A simplified ``no-go theorem''.}

These problems would go away if we had chirality, because it would
single out the $m=0$ values as special and remove the additive renormalization.
But, this cannot be done. There are rather deep reasons for why this
cannot be done, and several versions of ``no-go theorems'' exist\cite{nogo}. 
Here,
I shall present only a very simple version, namely, one cannot supplement
the above requirements of $D(m)$ (fulfilled by $D_W$) with the requirement
$\gamma_5 \tilde D \gamma_5 = -\tilde D$ at $m=0$. The proof is very simple:
Take the particular case where the $T_\mu$ commute. Compounding four
elementary reflections we get $\gamma_5 \tilde D(\theta ) \gamma_5 =
\tilde D (-\theta ) = - \tilde D(\theta )$. This shows that any one of
the sixteen solutions $\theta_\mu = 0,\pi$ has $\tilde D (\theta^* ) =0$,
since, by periodicity $\pi=-\pi$. This violation of one of the original
requirements amounts to a drastic multiplication of the number of
Dirac particles: instead of one we end up with sixteen !

\section{Overlap Dirac Operator}
Any regularization deemphasizes the dynamics of short length/time scales.
Lattice regularizations completely remove arbitrarily high momenta by
compactifying momentum space. Thus, the spectrum of a lattice regularized
Dirac operator lives on a compact space. When looking for a way around
the chirality ``no-go theorem'' a possible line of attack is to prepare
ahead of time for the regularization step by shifting the
focus to an operator which already has a compact spectrum in the continuum.

Based on the anti-hermiticity of $D_c$ (which we argued before can be viewed
as a fundamental property, one that in conjunction with the other fundamental
property of $\gamma_5$-hermiticity produces chirality as a consequence)
we can quite naturally define an associated unitary operator $V_c$
using the Cayley transform:
\begin{equation}
V_c = {{D_c -\Lambda_c }\over { D_c + \Lambda_c }} \enspace .
\end{equation}
$V_c$ is not only unitary, but also $\gamma_5$-hermitian, 
inheriting the property from $D_c$. In other words, $V_c$ has
the two crucial properties of $D_c$ and these two properties
would imply chirality for $D_c$ defined in terms of $V_c$ by
inverting the Cayley transform (the inverse is another Cayley transform).
Also, one has to note that the ``no-go theorem'' does not prohibit
a lattice version of $V_c$, $V$, that satisfies both requirements of 
$\gamma_5$-hermiticity and unitarity. However, the $D$ one would construct
from such a $V$ would need to violate something - the natural choice
is that $D$ be non-local. Still, nothing seems to say that $V$ itself
has to be nonlocal: 
\begin{equation}
D={{1+V}\over{1-V}} \enspace .
\end{equation}
The non-locality in $D$, mandated by the ``no-go theorem'', could
reflect merely the existence of unit eigenvalues to $V$. 

The next question is then, suppose we have a lattice $V$; since $D$
is non-local, how can we hope to make progress ? The answer is almost
trivial if we go back to continuum: we only care about the spectral
part of $D_c$ which is below some cutoff $\Lambda_c$. So, we are
allowed to expand the Cayley transform:
\begin{equation}
V_c = -1 +2{D_c\over \Lambda_c} +O\left ({D_c\over \Lambda_c}\right )^2
\enspace .
\end{equation}
Therefore,
\begin{equation}
{D_c \over \Lambda_c } = {{1+V_c}\over 2}~~~{\rm plus~unphysical~corrections.}
\end{equation}
If the lattice-$V$ is local there is no locality 
problem with the lattice-$D_o$
being given by
\begin{equation}
D_o = {{1+V}\over 2} \enspace .
\end{equation}
Now, in agreement with the ``no-go theorem'', we have lost exact 
anti-hermiticity, and, consequently exact, mathematical, 
chirality. However, the
loss of chirality can be made inconsequential, unlike in
the traditional Wilson formulation of fermions. As a first sign of
this let us check that adding a mass term produces a lower bound
similar to the continuum and thus would protect our precious condition
number. It is easy to prove that:
\begin{equation}
{\left ( {m_c\over \Lambda_c} +{{1+V_c }\over 2}\right )}^\dagger
{\left ( {m_c\over \Lambda_c} +{{1+V_c }\over 2}\right )}\ge {\rm min}
\left [ \left ({m_c\over \Lambda_c }\right )^2 , \left (
{m_c \over \Lambda_c } +1 \right )^2
\right ] \enspace .
\end{equation}
So, we have an expression bounded away from zero 
as long as ${m_c\over\Lambda_c} \neq  0, -1$. Another way to see
that we have as much chirality as we need is to calculate
the anticommutator of $D_o$ with $\gamma_5$, which would be zero
in the continuum, when $D_c$ is used. Actually, in the 
language of path integrals, we do not really need the Dirac
Operator itself: rather we need its inverse and its determinant.
So, it is closer to physics to look at the inverse of the Overlap
Dirac Operator $D_o$. $D_o^{-1}$  obeys
\begin{equation}
\gamma_5 D_o^{-1} + D_o^{-1}\gamma_5 = 2\gamma_5 \enspace .
\end{equation}
The same equation would also hold in the continuum if we replaced
$D_c^{-1}$ by ${2\over{1+V_c}}$. There is little doubt that such a
replacement in the continuum has no physical consequence; after all
it only changes $D_c^{-1}$ additively, by $-1$, 
a Dirac delta-function in space-time. 
While on the lattice we don't have a local and exactly chiral $D$, we
do have a $D_o={{1+V}\over 2}$ and we now see that {\it its} inverse is
good enough. 
In a brilliant paper \cite{gw}\footnote{
Clearly, when going to the lattice, the continuum 
delta-function can be replaced
by a Kroneker delta-function or by, say, a narrow Gaussian. This gives
a certain amount of freedom which I shall ignore below. In the lattice
formulation of Ginsparg and Wilson this freedom is made explicit by the
introduction of a local operator $R$.}, published almost twenty years
ago, Ginsparg and Wilson suggested  
this relaxation of the chirality condition as a way 
around the ``no-go theorem''.
What stopped progress was the failure
of these authors to also produce an explicit formula for $D_o$. Their
failure, apparently, was taken so seriously, that nobody tried to
find an explicit $D_o$, and 
the entire idea fell into oblivion for an embarrassingly long time. 
Nobody even made the relatively straightforward observation that the
search for $D_o$ was algebraically equivalent to a search for a 
unitary, $\gamma_5$-hermitian operator $V$. Moreover, it seems
to have gone under-appreciated that the problem
was not so much in satisfying the algebraic constraint of the relaxed
form of the anticommutator (the GW relation), but, in simultaneously
maintaining gauge covariance, discrete lattice symmetries, 
and correct low momentum behavior, without
extra particles.

\subsection{Definition and basic properties}

To guess a lattice formula for $V$, first notice that, in the continuum,
the operator $\epsilon_c = \gamma_5 V_c$ is a reflection: $\epsilon_c^2=1$
and $\epsilon_c^\dagger =\epsilon_c$. Also, $V_c$, and hence $\epsilon_c$,
are expressed as a ratio of {\it massive} Dirac operators. But, we
know that there is no difficulty in representing on the lattice the
massive Dirac Operator. The simplest way to do this is to use the
Wilson Dirac Operator, $D_W$. The latter obeys $\gamma_5$-hermiticity,
so $H_W = \gamma_5 D_W$ is hermitian. It is then very natural
to try
\begin{equation}
\epsilon={\rm sign} ( H_W ) \enspace .
\end {equation}

This is not the entire story - we still have one parameter at our
disposal, the lattice mass, $m$. When we look at the continuum
definition of $V_c$ we observe that it can be rewritten as follows:
\begin{equation}
V_c = {{D_c - \Lambda_c }\over \sqrt {-D_c^2 +\Lambda_c^2}}
       {\sqrt {-D_c^2 +\Lambda_c^2} \over {D_c + \Lambda_c }}
\enspace .
\end{equation}
So $V_c$ is given by the product of two unitary operators, one
being the unitary factor of the Dirac Operator with a large negative
mass and the other the conjugate of the 
unitary factor of the Dirac Operator with
a large positive mass. Formally, the unitary factors are equal to each
other up to sign in the continuum and one has:
\begin{equation}
V_c =- U_c^2 =-{ \left (
{{D_c - \Lambda_c }\over \sqrt {-D_c^2 +\Lambda_c^2}} \right )}^2;~~~~~
-D_c^2 +\Lambda_c^2 = (D_c - \Lambda_c )^\dagger (D_c - \Lambda_c )
\enspace .
\end{equation}
But, on the lattice, a local unitary operator $U$ replacing $U_c$ cannot
exist; if it did, $D={1\over 2}(U^\dagger -U)$ 
would violate the ``no-go theorem''.
So, on the lattice we can find a V, but there is no 
local square root of $-V$; this observation is rather deep, and related
to anomalies. We need to treat the two unitary factors in the continuum
expression for $V_c$ differently on the lattice. Actually, the factor
with a positive mass sign can be replaced by unity by taking the limit
$m\to\infty$, something one can 
do only on the lattice, where the $T_\mu$ operators 
are bounded. The factor representing negative mass however cannot be so
simplified because the negative mass argument $m$ is restricted to a finite
interval ($-2<m<0$) to avoid extra particles. Thus, on the lattice
we are led to 
\begin{equation}
V=D_W (m) {1\over {\sqrt{D_W^\dagger (m) D_W (m)}}}~~~~{\rm with}~~~-2<m<0
\enspace . 
\end{equation}
This formula is equivalent to the one for $\epsilon$ above.
The reason that we get exactly massless quarks is that no fine
tuning is needed for $V$ to have eigenvalues very close to $-1$.
Indeed, $D_W (m) = D_W (0) +m$ and, in the case $[T_\mu , T_\nu ]=0$,
$D_W(0)$ is easily seen to have very small eigenvalues. There $D_W(m)$
is dominated by the negative mass $m$. It is unimportant what the
exact value of $m$ is, only its sign matters. Even when $[T_\mu , T_\nu ]
\neq 0$, and $m$ is additively renormalized, as long 
as the effective $m$ stays negative, we shall have exactly 
massless quarks as evidenced by 
the eigenvalues of $V$ close to $-1$ potentially producing long range
correlations in $D_o^{-1}$. 

All non-real eigenvalues of $V$ are 
paired: $V\psi = e^{iv} \psi \Rightarrow 
V\gamma_5 \psi = e^{-iv} \gamma_5\psi$. But, one important property
of the continuum $V_c$ is that it has an unpaired single $-1$ eigenvalue
in a certain class of topologically nontrivial backgrounds. This is
a characteristic of massless quarks and has to be reproduced on the
lattice. There $V$ is even dimensional, so a single exact $V=-1$ 
state implies the existence of another $V=+1$ state. In addition,
it is clear that as a result of $|m|<2$, there are states where
$D_W (0)$ dominates over $m$ and these states will have eigenvalues
far off the real axis. We conclude that the spectrum of $V$ can cover
the entire complex unit circle. The spectrum of $U_c$ though, covers
only half the complex unit circle. This is a reflection of $V$ being
a lattice version of $-U_c^2$ rather than $U_c$ itself, as one
might think naively.

It was mentioned already that all one needs in the path integral are
formulae for the inverse of the Dirac Operator and for its determinant.
It is clear that once one has an acceptable $V$ one can form not only
the local, but not strictly chiral operator $D_o$ but also a non-local
chiral associate $D_o^\chi={{1+V}\over{1-V}}$. In the determinant one
must use $D_o$ and cannot use $D_o^\chi$. Otherwise, 
the extra zeros of the determinant coming from the $V=1$ states
leave an indelible effect in the continuum limit. These zeros directly
reflect the non-locality of $D_o^\chi$. However, the inverses of $D_o$
can be replaced by inverses of $D_o^\chi$ which has useful practical
implications. In the language of Feynman graphs, the determinant
is represented by closed internal fermion loops, and on those
we must use $D_o^{-1}$, but the inverses come from external fermions
lines, and on those we can just as well use $D_o^{\chi~-1}=D_o^{-1}-1$.
This is consistent because the $-1$ term can be interpreted 
as coming from an auxiliary fermionic variable which contributes only
a unit multiplicative factor to the fermionic determinant. 

To complete the logic of the story let me discuss the introduction of
a finite quark mass parameter in the context of the Overlap Dirac 
Operator. To be sure there is no additive mass renormalization, one
wishes to preserve the continuum property that $Tr {1\over {D_c + m_c}}$
is odd in $m_c$; this property\footnote{
A less careful introduction of non-zero quark mass 
in the Overlap Dirac Operator requires extra unnecessary
subtractions - see \cite{jansen}.} singles out the point $m_c =0$. This
property is a direct consequence of $\gamma_5 D_c + D_c \gamma_5 =0$.
It is easy to see that for an external fermion line the lattice
version would be
\begin{equation}
Tr {{e^\rho - V}\over {e^\rho +V}} = Tr \left ( {2\over{1+e^{-\rho} V}}
-1 \right );~~~~~\rho > 0;
\end{equation}
with 
\begin{equation}
m_q = z \tanh {\rho\over 2};~~~~\left ( {{1+V}\over 2}
\right )_{T_\mu = \exp (i\theta_\mu )}
\longrightarrow {i\over z} \sum_\mu \gamma_\mu \theta_\mu + O(\theta_\mu ^2 )
\enspace .
\end{equation}
The factor $z$ is somewhere between $1$ and $2$. One now easily proves that
the condition number of the matrix ${{e^\rho + V}\over {e^\rho - V}}$
is ${1\over {\tanh({\rho\over 2} )}}={z\over  m_q}$. 
Therefore, we should be able 
to get to the quark masses we need. However, the matrix $\epsilon$ isn't
sparse, and the evaluation of its action will be time consuming. Still, it
is a function of a sparse matrix $H_W$, so employing the Overlap Dirac
Operator in numerical QCD is not ruled out {\it ab initio}. 

\subsection{Implementation by rational approximants}

The operator $\epsilon$ is unambiguously defined only for matrices $H_W$
that have no zero eigenvalues. This is not a restriction in itself, because
the variables $U_\mu (x)$ in $H_W$ are stochastic and there is no symmetry
protecting zero eigenvalues of $H_W$. So, the probability to encounter
an exact zero of $H_W$ is zero. Moreover, in the continuum limit
$H_W$ will have a relatively large gap around zero, so when we are close
enough to the continuum we need the sign function only over the
range of arguments $[-a,-b] \cup [b,a]$, where $a\sim 8$ and $b$ should
be something like $0.1$ or $0.5$.

Over the above range it is relatively
easy to approximate the sign function by simpler functions. 
Also, for $b$ uniformly (in the background) bounded away from
zero, the operator ${1\over \sqrt{D_W^\dagger D_W}}$ is a convergent
series in $H_W$. This ensures that the non-sparse $D_o$ is 
nevertheless sufficiently local to be an acceptable
approximation to a continuum differential operator. 

Mimicking 
the way transcendental 
functions are typically implemented in computers, we are led
to try a rational approximation \cite{rational}. 
So, we are looking for a series of
functions $\varepsilon_n (x)$ which, for any fixed $x \neq 0$, obeys
$\lim_{n\to\infty} \varepsilon_n (x) = {\rm sign}(x)$. For each finite
$n$, $\varepsilon_n (x)$ is a ratio of two polynomials. 
A sequence with many good properties has been in use by 
applied mathematicians \cite{higham} 
who were interested in the sign function because
of the role it plays in control theory. The sequence is given by:
\begin{equation}
\varepsilon_n (x) = {{(1+x)^{2n} - (1-x)^{2n}}\over
{(1+x)^{2n} + (1-x)^{2n}}} = 
\cases{
|x|<1 &$\tanh[2n\tanh^{-1} (x)]$\cr |x|>1
&$\tanh[2n\tanh^{-1}(x^{-1})]$\cr
|x|=1&$x$\cr}
\end{equation}
So long $|x|$ is sufficiently far from zero a large enough $n$
can provide an approximation to the sign function good to machine
accuracy. The approximants $\varepsilon_n (x)$ are smooth and 
maintain some properties of the sign function:
\begin{equation}
\varepsilon_n (x) = - \varepsilon_n (-x)=\varepsilon_n ({1\over x});~~~~~~
|\varepsilon_n (x)| \le 1;~~~~~\varepsilon_n (\pm 1)=\pm 1
\enspace .
\end{equation}

To implement $\varepsilon_n (x)$ we decompose $f(x^2)=\varepsilon_n (x)/x$
in simple pole terms:
\begin{equation}
\varepsilon_n (x)={x\over n} \sum_{s=1}^n {1\over{
x^2 \cos^2 {\pi\over{2n}} (s-{1\over 2})+\sin^2 
{\pi\over {2n}} (s-{1\over 2})}}
\enspace .
\end{equation}
One can choose an appropriate scaling parameter $\lambda > 0$ and approximate
the sign function of $H_W$ by $\varepsilon_n (\lambda H_W )$. The action
of this approximated $\epsilon$ on a vector $\psi$ can be evaluated by 
a multi-shift Conjugate Gradient inversion algorithm \cite{multishift}. 
The multi-shift
trick reduces the computational cost of evaluating the action of each
one of the terms in the pole expansion to the cost of evaluating one
single inversion, the most time consuming one, which clearly is
\begin{equation}
{1\over{\lambda^2 H_W^2 + \tan^2 {\pi\over {4n}}}}\psi \enspace .
\end{equation}
Memory requirements are linear in $n$, since one needs to store
a few vectors for each pole term. In practice this may be a problem,
not as much that the entire memory of the machine would be exceeded,
but rather that one would find oneself exceeding the level 2 cache.
Level 2 cache misses can lead to substantial performance loss. Thus,
it is useful to consider an alternative to the standard implementation
of the multi-shift trick. Usually, in applications 
of the multi-shift trick, one is really interested in the
individual vectors, but here we only want their weighted sum. It is quite
easy to figure out a way to store only a few vectors, and get the sum, but
a double pass over the basic Conjugate Gradient procedure
is now required. Although the number of floating point operations is doubled,
because of reduced cache miss penalties, performance is not
necessarily adversely impacted and can actually increase \cite{twopass}.

Other promising methods to implement $D_o$ have been developed and
will be hopefully discussed here later \cite{borici}. In the implementation
employed in \cite{jansen} the sign function was approximated
by a very high order polynomial. 
 
\subsection{A new bottleneck and projectors}
Having removed one apparent obstacle, namely the potentially
large memory requirements, we turn to a much more substantial
obstacle, namely the condition number of $H_W^2$, $\kappa$.
$\kappa$ determines both the number of Conjugate Gradient iterations
needed to evaluate the inverse and also the minimal required $n$
for given expected accuracy $\delta$, where
\begin{equation}
\|\varepsilon_n (\lambda H_W )-{\rm sign} (H_W )\|< \delta
\enspace .
\end{equation}
The optimal choice of $\lambda$ is easily found because of the
inversion symmetry of $\varepsilon_n (x)$: $\lambda h_{\rm min} =
{1\over{\lambda h_{\rm max}}}$ where $h_{\rm min,~max}$ are the
square roots of the minimal and maximal eigenvalues of $H_W^2$. 
Thus, $\lambda={1\over \sqrt{h_{\rm min} h_{\rm max}}}$ so the
range over which the sign function is needed is 
\begin{equation}
{1\over \kappa^{1\over 4}} < |x| < \kappa^{1\over 4}\enspace .
\end{equation}
For $\varepsilon_n (x)$ to be an acceptable approximation we need then
$n\approx {1\over 4} \kappa^{1\over 4} |\log (\delta / 2 )|$. For example,
for $\kappa \sim 10^4$ and single precision $n\sim 50$ is safe; for
double precision, the needed $n$ doubles. 
On the other hand the number of Conjugate Gradient iterations will go as
$\kappa^{1\over 2}$. Note that the lowest eigenvalue of 
$\lambda ^2 H_W^2 $ and the minimal pole shift ${\pi^2 \over {16n^2}}$
are of the same order when $n\sim \kappa^{1\over 4}$.

Comparing to traditional simulations, where $\kappa$ is again 
the source of all problems, it is important to note
the new feature that now the parameter $m$ in $H_W$
is taken in a different regime from before. In traditional simulations 
the parameter $m$ is adjusted so that the physical quark mass be small.
This precisely means that the parameter is chosen so that $\kappa$
be large. Because of fluctuations, $\kappa$ becomes often much larger
than one would have expected, and that is the problem faced by
traditional simulations discussed earlier. Here the parameter $m$,
subject to the limitation $-2< m <0 $, can be chosen so that 
$\kappa$ be as small as possible. Fluctuations can still cause problems
and sometimes make $\kappa$ large, but, in principle, the fluctuations
are around a small $\kappa$ value, not a large one. In practice however
the situation is somewhat marginal: coarse lattices produce too much
fluctuations even for the Overlap Dirac Operator. But, finer lattices
are manageable. Still, even on finer lattices one needs to split the domain
over which the sign function is approximated into a neighborhood around
zero and the rest. The neighborhood around zero contains of the order of 
ten eigenvalues, and by computing the exact projector on the corresponding
eigenspace, the sign function is exactly evaluated there, leaving only
the more manageable part of the spectrum to be dealt with by the rational
approximation \cite{scri}. 
This is quite time consuming, and constitutes the new bottleneck 
we are facing. Variations of the probability distribution of the background
and slight modifications of $H_W$ would not affect the continuum limit
but would probably help ameliorate this problem.
\subsection{Avoiding nested Conjugate Gradient procedures}

Another obvious drawback of using the Overlap Dirac Operator in numerical
simulations is that we eventually need to compute vectors of the form
${2\over{1+V}} \psi$ for a given $\psi$. 
This requires another inversion and we end up with
a two level nested Conjugate Gradient procedure, whereas in the traditional
simulations we only had one. The outer Conjugate Gradient is going to be
better conditioned (for light but still massive quarks) then in the 
traditional approach. So, we are facing a tradeoff issue, one that has
not been resolved yet. The two levels of Conjugate Gradients are presently
dealt with separately, but clearly, eventually, potential 
gains will be obtained from the fact that the inner 
procedure does not have to be run to high accuracy for the initial
steps in the outer procedure. 

There exists another trick \cite{chains} to simulate
the rational $\varepsilon_n (\lambda H_W )$ which does away with
the need of employing a nested conjugate gradient altogether, but
at the expense of memory usage linear in $n$. Basically, the point is
to find a Conjugate Gradient algorithm operating in an enlarged space
and producing the vector ${2\over{1+V}} \psi$ in one blow. This trick is
based on two observations:
\begin{enumerate}
\item Any rational approximation can be written as a truncated 
continued fraction.
\item Any truncated continued fraction can be represented by a Gaussian
Path Integral of a fermionic system living on a chain of length $n$,
where $n$ is the depth of the truncated continued fraction.
\end{enumerate}

This method is applicable to any rational approximation. Below is one way
to apply it to $\varepsilon_n (H_W)$, where the scale factor is absorbed
in $H_W$ for notational simplicity.

First, the rational approximation
has to be written in the form of a continued fraction
with entries preferably linear in $H_W$. 
I start from a formula that goes as far back as Euler, 
and subsequently
use the invariance under inversion of $x$ to move the $x$ factors
around, so that the entries become linear in $x$.   
\begin{eqnarray}
&\varepsilon_n (x) ={ {\displaystyle 2nx }\over \displaystyle 1 +
                 {\strut {\displaystyle (4n^2-1)x^2 } 
\over\displaystyle 
                                      3+
                   {\strut {\displaystyle (4n^2-4)x^2 } 
\over\displaystyle 
                                      5 + \dots
                   {\strut \displaystyle   \ddots \over 
{\displaystyle 4n-3}+
                     {\strut {\displaystyle 
[4n^2 - (2n-1)^2]x^2} \over 
{\displaystyle 4n-1} }}}}}
\end{eqnarray}
Now, with the help of extra fields, I write a Gaussian path integral
which induces the desired action between a chosen subset of fields:
\begin{equation}
\int d\bar\phi_1 d\phi_1  d\bar\phi_2 d\phi_2 \dots  
d\bar\phi_n d\phi_n  e^{S_*} =
(\det H_W )^{2n} 
 e^{-\bar\psi (\gamma_5 +\varepsilon_n (H_W ) )\psi } \enspace .
\end{equation}
The quadratic action $S_*$ couples the 
extended fermionic fields $\bar \chi , \chi$:
\begin{equation}
\bar \chi = \pmatrix {\bar\psi & \bar\phi_1 & \dots & \bar\phi_{2n}\cr},~~
\chi  =  \pmatrix {\psi & \cr\phi_1 & \cr\vdots &\cr \phi_{2n}} \enspace .
\end{equation}
$S_* = \bar\chi {\bf H} \chi$, where the new kernel, ${\bf H}$, 
has the following block structure:
\begin{equation}
{\bf H}=\pmatrix{-\gamma_5 & \sqrt {\alpha_0 }& 0&  \dots &\dots & 0 \cr
          \sqrt {\alpha_0 }& H_W & \sqrt {\alpha_1 }& \dots &\dots &0\cr
           0& \sqrt {\alpha_1 } & -H_W &  \dots&\dots & 0\cr
                  \dots & \dots & \dots & \ddots & \dots & 0\cr
                   \dots & \dots & \dots & \dots & H_W & 
                                            \sqrt{\alpha_{2n-1}}\cr
	          \dots & \dots & \dots & \dots & \sqrt{\alpha_{2n-1}}
& -H_W \cr}
\end{equation}
The numerical coefficients $\alpha$ are given below:
\begin{equation}
\alpha_0 =2n ,~
\alpha_j = {{(2n-j)(2n+j)}\over {(2j-1)(2j+1)}},~ j=1,2,...
\end{equation}
The hope is that the condition number of ${\bf H}$ will be manageable. 
The basic point is that up to a scalar factor, the (1,1) diagonal
block of the inverse ${\bf H}^{-1}$ is equal to ${1\over {1+V}}\gamma_5$.
${\bf H}$ is sparse and the evaluation of ${\bf H}^{-1} \chi$ requires
one single Conjugate Gradient procedure, albeit one acting on vectors
$2n$ times longer than needed. Although I used Path Integrals to get
the relation between ${\bf H}$ and the Overlap Dirac Operator, there
is nothing more to the derivation than ordinary Linear Algebra, and
Path Integrals could have been bypassed altogether.

So, at the expense of adding extra fields one can avoid
a nested Conjugate Gradient procedure. This would
be particularly important when dynamical fermions are simulated.
The chain version of the direct truncation of the Overlap
Dirac Operator is similar in appearance to 
another truncation, usually referred to as ``domain walls'' \cite{dmf}.

The proposal above, employing chains, has two potential
advantages over domain walls. First, it is much more flexible,
allowing one to change both the rational approximation
one uses and its chain implementation. This flexibility
ought to allow various improvements. 
Second, 
since here the argument of the approximated sign function is $\lambda H_W$, not
the rather cumbersome logarithm of the transfer matrix of the domain wall
case, eigenstates of $H_W$ with small eigenvalues  can be eliminated by
projection with greater ease. This elimination, although costly
numerically, vastly increases the accuracy of the approximation to the sign
function. Actually, at this stage of the game and at practical gauge
coupling values, the use of projectors seems to be numerically indispensable
to direct implementations of the Overlap Dirac Operator. 
Although no projectors have been implemented in domain wall
simulations and physical results have been claimed, to me it seems doubtful
that the domain wall version of truncating the Overlap 
will really be capable to get to small
quark masses without employing some technique equivalent to projection. 
\section{Acknowledgments}
I am grateful to the organizers of the Interdisciplinary Workshop
on Numerical Challenges to Lattice QCD for the opportunity to participate
and to describe the Overlap. I wish to thank them for their 
warm hospitality. My research at Rutgers is supported by the DOE under grant
\# DE-FG05-96ER40559.

\end{document}